\documentclass[12pt]{article}
\usepackage{sw20lart}

\input tcilatex
\QQQ{Language}{
British English
}

\begin{document}

\title{Evolving spheres of shear-free anisotropic fluid}
\author{B.V.Ivanov \\
Institute for Nuclear Research and Nuclear Energy, \\
Bulgarian Academy of Science, \\
Tzarigradsko Shausse 72, Sofia 1784, Bulgaria}
\maketitle

\begin{abstract}
The fluid models mentioned in the title are studied in a modified approach,
based on two formulas for the mass function. All characteristics of the
fluid are expressed through a master potential, satisfying an ordinary
second order differential equation. Different constraints are imposed on
this core of relations, finding new solutions and deriving the classical
results for perfect fluids as particular cases. All charged anisotropic
solutions, all conformally flat and all uniform density solutions are found.
A large class of solutions with linear equation among the two pressures is
derived, including the case of vanishing tangential pressure. The mechanism
responsible for the appearance of equation of state is elucidated.
\end{abstract}

\section{Introduction}

The description of gravitational collapse and evolution of compact objects
under various conditions remain among the important problems of general
relativity. They are described by spherically symmetric relativistic fluid
models where the metric depends on time and radius. In general, they possess
shear, expansion and acceleration, which makes them rather hard to study. In
many physical situations, however, shear-free perfect fluid models are a
good approximation. The history of their study is long and rich in
rediscoveries \cite{one}. McVittie used an ad hoc ansatz to find a solution 
\cite{two}. The solution of the Einstein equations may be reduced to a
single ordinary second order differential equation \cite{three}, \cite{four}%
. Its integration may be done by Lie point symmetry methods \cite{four}, 
\cite{five}, \cite{six}, \cite{seven}, \cite{eight}. Another approach is to
search for solutions which have the Painlev\'e property \cite{nine}. Other
solutions have been found too \cite{ten}.

The study of charged shear-free perfect fluids started with the papers of
Shah and Vaidya \cite{eleven} and Faulkes \cite{twelve}. The master equation
acquires an additional non-derivative term. The three approaches mentioned
above were applied to the charged case too finding new solutions \cite
{thirteen}, \cite{fourteen}, \cite{fifteen}, \cite{sixteen}, \cite{seventeen}%
, \cite{eighteen}, \cite{nineteen}, \cite{twenty}, \cite{twone}. It was
noticed that both in the neutral and in the charged case the second Weyl
invariant is tightly bound to the arbitrary functions in the master
equation. Solutions which obey an equation of state (EOS) have been
discussed in the neutral \cite{three}, \cite{twtwo} and the charged case 
\cite{twthree}, \cite{twfour}, \cite{twfive}, \cite{twsix} and are known as
the Wyman and the charged Wyman solution.

All perfect fluid solutions with uniform energy density have been found by
Kustaanheimo \cite{one}, \cite{twseven}. There are two branches - one in
elementary functions and one given implicitly. Different properties of these
solutions have been discussed \cite{tweight}, \cite{twnine}, \cite{thirty}, 
\cite{thione}, \cite{thitwo}, \cite{thithree}, \cite{sixteen}, \cite{thifour}%
. In some of these papers it was noticed that conformally flat solutions are
particular cases of uniform density solutions. Charged shear-free perfect
fluid spheres with uniform density are necessarily static \cite{thifive}. As
for the conformally flat perfect fluid solutions, all of them have been
found by Stephani, using an embedding in a flat sixdimensional space \cite
{one}, \cite{thisix}. Some approximate conformally flat solutions for
anisotropic fluid with slow motion were studied recently \cite{thiseven}, as
well as conformally flat solutions for radiating perfect fluids \cite
{thieight}, \cite{thinine}. Close to them are solutions admitting conformal
symmetry \cite{one}, \cite{forty}, \cite{foone}.

On the other hand, different mechanisms have been identified through the
years which create pressure anisotropy in stellar models and make the fluid
imperfect \cite{fotwo}. Such are the exotic phase transitions during
gravitational collapse, the existence of a solid core or the presence of a
type P superfluid. Viscosity may also be a source of local anisotropy as
well as the slow rotation of a fluid. It has been shown that the sum of two
perfect fluids, two null fluids or a perfect and a null fluid may be
represented by effective anisotropic fluid models \cite{fothree}. Recently
it was pointed out that the same is true for perfect fluids with charge,
bulk and shear viscosity \cite{fofour}.

The aim of the present paper is to generalize the above results, valid for
shear-free perfect fluids, to shear-free anisotropic fluids. For this
purpose we use the reformulation of the Einstein equations in terms of the
mass function \cite{fofive}, \cite{fosix}. It is combined with a formula for
the same function derived recently \cite{fotwo}, \cite{foseven}, \cite
{foeight} and adapted to our goals. In this approach, besides the new
solutions, the classical results reviewed above follow as particular cases
when the isotropy condition is imposed.

In Sec.2 we give a derivation of the mass decomposition formula in the
general case of radiating spherically symmetric anisotropic fluid when both
heat flow and null fluid are present. It consists of a local and a global
part, the latter being the second Weyl invariant. In Sec.3 the field
equations based on the mass function are given for non-radiating spherically
symmetric metric. The fundamental potential $Z$ is introduced and a new
formula for the shear is presented. Sec.4 discusses the general solution of
the shear-free anisotropic model based on a single second order differential
equation. In Sec.5 the particular cases of perfect fluids and charged
anisotropic fluids are solved, making contact with the existing literature.
In Sec.6 we impose in addition to the isotropy condition the requirement for
EOS and rederive the Wyman and the charged Wyman solution. In Sec.7 the
general conformally flat solution for anisotropic shear-free fluids is found
and the classic Stephani solution is obtained in particular. The general
uniform density solution is derived in Sec.8 and the known solution for
perfect fluid is reobtained. A mechanism of producing models with special
linear EOS is outlined in Sec.9. Sec.10 studies the case when the radial and
the tangential pressures are proportional or one of them vanishes. Finally,
some conclusions are drawn in the last section.

\section{Mass decomposition formula}

Spherically symmetric relativistic fluid models are described by the metric 
\begin{equation}
ds^2=e^{2\nu }dt^2-e^{2\lambda }dr^2-Y^2\left( d\theta ^2+\sin ^2\theta
d\varphi ^2\right) ,  \label{one}
\end{equation}
where $\nu ,\lambda $ and $Y$ are functions of $t$ and $r$ only. The
spherical coordinates are numbered as $x^0=t,x^1=r,x^2=\theta ,x^3=\varphi $%
. In general relativity one defines the Weyl tensor $C_{ijkl}$ through the
Riemann tensor of the spacetime $R_{ijkl}$ as follows \cite{fonine} 
\begin{equation}
R_{ijkl}=C_{ijkl}+\frac 12\left(
g_{ik}R_{jl}+g_{jl}R_{ik}-g_{jk}R_{il}-g_{il}R_{jk}\right) -\frac 16\left(
g_{ik}g_{jl}-g_{jk}g_{il}\right) R.  \label{two}
\end{equation}
Here $R_{ij}$ is the Ricci tensor and $R$ is its trace, the Ricci scalar.
The mass function $m$ is given by 
\begin{equation}
m=\frac 12YR_{\ 232}^3.  \label{three}
\end{equation}
We need the Einstein equations 
\begin{equation}
R_{ij}-\frac 12g_{ij}R=8\pi T_{ij},  \label{four}
\end{equation}
where $T_{ij}$ is the energy-momentum tensor (EMT) and we have set $G=c=1$.
For an anisotropic fluid model with heat flow one has \cite{fifty} 
\begin{equation}
T_{ik}=\left( \rho +p_t\right) u_iu_k-p_tg_{ik}+\left( p_r-p_t\right) \chi
_i\chi _k+q_iu_k+q_ku_i.  \label{five}
\end{equation}
Here $\rho $ is the energy density, $p_r$ is the radial pressure, $p_t$ is
the tangential pressure, $u^i$ is the four-velocity of the fluid, $\chi ^i$
is a unit spacelike vector along the radial direction and $q^i$ is the heat
flux (in the radial direction too). They satisfy the relations 
\begin{equation}
u^iu_i=1,\qquad \chi ^i\chi _i=-1,\qquad u^i\chi _i=0,\qquad u^iq_i=0.
\label{six}
\end{equation}
The coordinates are assumed to be comoving, hence 
\begin{equation}
u^i=e^{-\nu }\delta _0^i,\qquad \chi ^i=e^{-\lambda }\delta _1^i,\qquad
q^i=qe^{-\lambda }\delta _1^i,  \label{seven}
\end{equation}
where $q=q\left( r,t\right) $. The traces of Eqs (4,5) give respectively 
\begin{equation}
R=-8\pi T,\qquad T=\rho -2p_t-p_r,  \label{eight}
\end{equation}
$T$ being the trace of the EMT.

Now we are in position to rewrite the basic Eq (2) in terms of $T_{ij}$
instead of $R_{ij}$%
\begin{equation}
R_{ijkl}=C_{ijkl}-4\pi \left(
g_{jk}T_{il}+g_{il}T_{jk}-g_{ik}T_{jl}-g_{jl}T_{ik}\right) -\frac{8\pi }%
3\left( g_{ik}g_{jl}-g_{jk}g_{il}\right) T.  \label{nine}
\end{equation}
Let us note that for non-trivial $u_i,\chi _{i\text{ }}$ and $q_i,$ $i=0$ or 
$1.$ Then for $i,j,k,l$ equal to $2$ or $3$ the previous equation greatly
simplifies 
\begin{equation}
R_{ijkl}=C_{ijkl}-\frac{8\pi }3\left( \rho +p_t-p_r\right) \left(
g_{ik}g_{jl}-g_{jk}g_{il}\right) .  \label{ten}
\end{equation}
This gives for the mass function 
\begin{equation}
m=\frac{4\pi }3Y^3\left( \rho +p_t-p_r\right) +\frac Y2C_{\ 232}^3.
\label{eleven}
\end{equation}
Note that the tangential pressure enters this formula thanks to the trace $T$%
, while the heat flow coefficient $q$ does not appear at all. A similar
formula has been found by inspecting the field equations \cite{foseven},
where non-comoving system of coordinates has been used and the expression in
the brackets reads $T_0^0+T_1^1-T_2^2$. In non-comoving coordinates the
individual components depend also on the radial velocity of the moving
observer. One can check, however, that in the cited combination this
dependence is cancelled and it coincides with the bracket term from Eq (11).
We have proved this formula without resorting to the explicit expressions
for the field equations. It is consequence of the 'orthogonal' character of
the four-vectors present in the EMT (pointing in the time and radial
direction) with respect to $m,$ defined in terms of tensor components along
the two spherical angles $\theta $ and $\varphi $.

Formula (11) holds also when the basic anisotropic fluid radiates a null
fluid with EMT 
\begin{equation}
T_{ij}^N=\varepsilon l_il_j,  \label{twelve}
\end{equation}
where $l^i$ is a null vector, $l^i=u^i+\chi ^i$ . One can show \cite{fofour}
that it adds to the basic $T_{ik}$ effective energy density, radial pressure
and heat flow, all of them equal

\begin{equation}
\rho ^N=p_r^N=q^N=\varepsilon .  \label{thirteen}
\end{equation}
This addition does not change Eq (11) or, put in other words, $\varepsilon $
(like $q$) does not appear in it directly.

Next we shall use the definition of the second Weyl invariant $\Psi _2$ in
terms of the Weyl tensor for the spherically symmetric case \cite{fione}. It
reads after a slight transformation 
\begin{equation}
2\Psi _2=\frac{C_{2323}}{g_{22}g_{33}}=-Y^{-2}C_{\ 232}^3.  \label{fourteen}
\end{equation}
Hence, one can rewrite formula (11) as 
\begin{equation}
m=\frac{4\pi }3Y^3\left( \rho +p_t-p_r\right) -Y^3\Psi _2.  \label{fifteen}
\end{equation}
In the static case one can set $Y=r$. For a perfect fluid $m$ contains only
its energy density. In the general case the anisotropic factor $\Delta
p=p_t-p_r$ is added. Thus, the total mass function is a sum of the local
part arising directly from matter sources and the Weyl term, responsible for
some global curvature. This last term may be related also to the electric
part of the Weyl tensor \cite{foeight}

\section{Field equations}

From now on we accept that there is no radiation ($q=\varepsilon =0$). We
also change the signature of the metric. The usual field equations (4) are
rather cumbersome except for the (01) component 
\begin{equation}
\dot Y^{\prime }-\dot Y\nu ^{\prime }-Y^{\prime }\dot \lambda =0.
\label{sixteen}
\end{equation}
The dot above means time derivative, while the prime denotes a radial one.
In order to make use of the mass decomposition formula (15) we shall work in
the formalism based on $m$ \cite{fofive}, \cite{fosix}, namely 
\begin{equation}
m^{\prime }=4\pi \rho Y^2Y^{\prime },  \label{seventeen}
\end{equation}
\begin{equation}
\dot m=-4\pi p_rY^2\dot Y,  \label{eighteen}
\end{equation}
\begin{equation}
\dot \lambda \left( \rho +p_r\right) =-\dot \rho -\frac{2\dot Y}Y\left( \rho
+p_t\right) ,  \label{nineteen}
\end{equation}
\begin{equation}
\nu ^{\prime }\left( \rho +p_r\right) =-p_r^{\prime }+\frac{2Y^{\prime }}%
Y\left( p_t-p_r\right) .  \label{tw}
\end{equation}
The mass is given by Eq (3) which yields 
\begin{equation}
m=\frac 12Y\left( 1+e^{-2\nu }\dot Y^2-e^{-2\lambda }Y^{\prime 2}\right) .
\label{twone}
\end{equation}
The expansion of the fluid is 
\begin{equation}
\Theta =e^{-\nu }\left( \dot \lambda +\frac{2\dot Y}Y\right) ,  \label{twtwo}
\end{equation}
while the components of the shear tensor are proportional to 
\begin{equation}
\sigma =\frac 13e^{-\nu }\left( \frac{\dot Y}Y-\dot \lambda \right) .
\label{twthree}
\end{equation}

Let us introduce next a quantity $Z$, which is central in the following 
\begin{equation}
Z=\Psi _2-\frac{4\pi }3\Delta p.  \label{twfour}
\end{equation}
Then the mass decomposition formula (15) may be written as 
\begin{equation}
\frac{4\pi }3\rho =\frac m{Y^3}+Z.  \label{twfive}
\end{equation}
This is an expression for the energy density which, unlike Eq (17), contains
the mass and not its derivative. Making use of Eq (17) we obtain 
\begin{equation}
m-\frac{m^{\prime }Y}{3Y^{\prime }}=-Y^3Z  \label{twsix}
\end{equation}
or in a more compact form 
\begin{equation}
\left( \frac m{Y^3}\right) ^{\prime }\frac Y{3Y^{\prime }}=Z.
\label{twseven}
\end{equation}

Next we find another expression for the shear. Replacing $m$ from Eq (15)
into Eq (18) gives 
\begin{equation}
3\dot Y\left( \rho +p_t\right) =Y\dot p_r-Y\left( \dot \rho +\dot p_t\right)
+\frac{3\left( Y^3\Psi _2\right) ^{.}}{4\pi Y^2}.  \label{tweight}
\end{equation}
Replacing in Eq (23) $\dot \lambda $ from Eq (19) and using the previous
equation we find 
\begin{equation}
\left( \rho +p_r\right) \sigma =\frac{e^{-\nu }}{3Y}\left[ \frac{3\left(
Y^3\Psi _2\right) ^{.}}{4\pi Y^2}-\left( Y\Delta p\right) ^{.}\right] .
\label{twnine}
\end{equation}
The formulae in this section hold for any spherically symmetric metric and
no radiation.

\section{Shear-free anisotropic fluids}

Let us impose on the metric the shear-free condition $\sigma =0$. Eq (23)
can be integrated up to an arbitrary function of $r$. Usually it is chosen
in such a way that 
\begin{equation}
e^{2\lambda }=\frac{Y^2}{r^2}.  \label{thi}
\end{equation}
Then Eq (16) is also integrable and determines $\nu $ up to an arbitrary
function of time. Eq (22) shows that it is related to the expansion of the
fluid and finally 
\begin{equation}
e^{2\nu }=\frac 9{\Theta ^2}\frac{\dot Y^2}{Y^2},  \label{thione}
\end{equation}
where $\Theta \left( t\right) $ is an arbitrary function \cite{one}.

In the following we shall use sometimes the notation 
\begin{equation}
L=\frac rY,\qquad x=r^2.  \label{thitwo}
\end{equation}
The metric may be written in terms of $Y$ or $L$%
\begin{equation}
ds^2=-\frac 9{\Theta ^2}\frac{\dot Y^2}{Y^2}dt^2+\frac{Y^2}{r^2}%
dr^2+Y^2d\Omega ^2=L^{-2}\left( -\frac 9{\Theta ^2}\dot
L^2dt^2+dr^2+r^2d\Omega ^2\right) .  \label{thithree}
\end{equation}
Here $d\Omega $ denotes the angular part. The same is true for Eq (21) 
\begin{equation}
m=\frac 12Y\left( 1+\frac{\Theta ^2}9Y^2-\frac{r^2Y^{\prime 2}}{Y^2}\right)
=Y^3\left[ \frac{\Theta ^2}{18}+2L_x\left( L-xL_x\right) \right] ,
\label{thifour}
\end{equation}
which leads to a formula for the density 
\begin{equation}
\frac{4\pi }3\rho =\frac{\Theta ^2}{18}+2L_x\left( L-xL_x\right) +Z.
\label{thifive}
\end{equation}
Eq (27) simplifies considerably because of Eq (34) and becomes 
\begin{equation}
\frac 43xLL_{xx}=Z.  \label{thisix}
\end{equation}

Let us utilize next the second formula for the shear, namely Eq (29). It
supplies the equation 
\begin{equation}
\frac{3\left( Y^3\Psi _2\right) ^{.}}{4\pi Y^2}-\left( Y\Delta p\right)
^{.}=0=\left( \frac{\Psi _2}{L^3}\right) ^{.}L^2-\frac{4\pi }3\left( \frac{%
\Delta p}L\right) ^{.}.  \label{thiseven}
\end{equation}
Combining it with the definition of $Z$ one expresses both $\Delta p$ and $%
\Psi _2$ through $Z$%
\begin{equation}
\frac{4\pi }3\Delta p=-\frac{\left( Y^3Z\right) ^{.}}{2Y^2\dot Y}=\frac{L^4}{%
2\dot L}\left( \frac Z{L^3}\right) ^{.},  \label{thieight}
\end{equation}
\begin{equation}
\Psi _2=-\frac{\left( YZ\right) ^{.}}{2\dot Y}=\frac{L^2}{2\dot L}\left(
\frac ZL\right) ^{.}.  \label{thinine}
\end{equation}
The radial pressure is determined from Eq (18) where the mass is given by Eq
(34). Together with the above formula for the anisotropy factor and Eq (25)
for the energy density they give two expressions for the tangential pressure 
\begin{equation}
4\pi p_tY^2\dot Y=-\left( m+\frac 32Y^3Z\right) ^{.}=\frac 12\left( m-4\pi
\rho Y^3\right) ^{.}.  \label{fo}
\end{equation}
In the general formalism $\rho $ and $p_r$ are expressed through the mass by
the similar formulae (17,18) while $p_t$ should be found either from Eq (19)
or Eq (20). For shear-free fluids, however, it is given by the above
formula, similar in structure to the formula for the radial pressure.

Now we are in position to outline in two different ways the general solution
for shear-free anisotropic fluids. First, we choose an arbitrary $L\left(
r,t\right) $ and find the metric, the mass function and $Z$ according to Eqs
(33,34,36). Then the energy density, the pressures and the second Weyl
invariant are given correspondingly by Eqs (35,18,40,39). In the second way
we take an arbitrary $Z\left( r,t\right) $ and solve for $L$ from Eq (36),
promoting the integration constants to arbitrary functions of time. This,
however, is rather difficult. For example, if $Z$ is a power $x^{n+1}$, we
get an Emden-Fowler equation which has explicit solutions for $n=0,-1,-2$
only \cite{fitwo}. Different additional constraints on $Z$ will be discussed
in the following, leading to other solutions.

The shear-free condition (37) can be written in integral form too 
\begin{equation}
Z=-\frac{8\pi }{3Y^3}\int \Delta pY^2\dot Ydt+\frac{g\left( r\right) }{Y^3}=%
\frac{8\pi L^3}3\int \Delta p\frac{\dot L}{L^4}dt+h\left( x\right) L^3
\label{foone}
\end{equation}
where $g,$ $h$ are arbitrary functions of $r$ or $x$.

\section{Perfect fluids. Charging a fluid}

The arbitrary function in the general shear-free anisotropic fluid solution
indicates that we are free to impose another condition. This can be done in
a number of ways. As a first example we impose the isotropy condition $%
p_r=p_t$ which makes the fluid perfect. Hence, perfect fluid can be viewed
upon as an anisotropic fluid with linear EOS among the pressures. Then the
integral term in Eq (41) disappears and the central equation (36) becomes 
\begin{equation}
L_{xx}=F\left( x\right) L^2,\qquad F\left( x\right) =\frac{3h\left( x\right) 
}{4x}.  \label{fotwo}
\end{equation}
This equation is well-known \cite{three}, \cite{four}. When $F=\left(
x-a\right) ^n$ , where $a,n$ are constants, it becomes an Emden-Fowler
equation and possesses five analytical solutions for $n=0,-15/7,-20/7,-5/2,$ 
$-5$ \cite{nine}, \cite{fitwo}. All of them may be expressed through the
Weierstrass elliptic function. The known solutions of Eq (42) \cite{one}, 
\cite{nine}, \cite{ten} include some others, like the second Painlev\'e
transcendent. The time dependence appears in $L$ by turning the integration
constants into integration functions of time. The general formulae give
expressions for the other characteristics of the fluid and the freedom in
the solution is reduced to arbitrary functions of time. Eq (24) shows that
for $\Delta p=0$ we have $Z=\Psi _2$ and hence 
\begin{equation}
\Psi _2=\frac 43xF\left( x\right) L^3,  \label{fothree}
\end{equation}
which too is a classical result \cite{one}, \cite{sixteen}. We see that the
appearance of the second Weyl invariant is due to the mass decomposition
formula, which in its turn stems from the definition of the Weyl tensor,
applied to an interior fluid solution.

Another way to constrain the form of $\Delta p$ is to charge the fluid.
Spherical symmetry allows the appearance of only an electric field $E$ in
the radial direction. The energy-momentum tensor of this field can be
described \cite{fofour} as addition of effective energy density and
pressures to the fluid with 
\begin{equation}
\rho ^E=p_t^E=-p_r^E=\frac{E^2}{8\pi }.  \label{fofour}
\end{equation}
The Maxwell equations give 
\begin{equation}
\Delta p^E=\frac{E^2}{4\pi }=f\left( x\right) L^4,\qquad 4\pi \tau
=Ee^{-\lambda },  \label{fofive}
\end{equation}
where $\tau $ is the charge function of the fluid and $f\left( x\right) $ is
an arbitrary function. Hence, Eq (41) acquires a charge term in addition to
other sources of anisotropy 
\begin{equation}
Z=h\left( x\right) L^3+\frac{8\pi }3f\left( x\right) L^4+\frac{8\pi }%
3L^3\int \Delta p\frac{\dot L}{L^4}dt.  \label{fosix}
\end{equation}
In the case of charged perfect fluid Eq (36) becomes 
\begin{equation}
L_{xx}=\frac{3h}{4x}L^2+\frac{2\pi f}xL^3,  \label{foseven}
\end{equation}
which is also well-known and studied in the past \cite{eleven}, \cite{twelve}%
. The Weyl invariant reads 
\begin{equation}
\Psi _2=hL^3+4\pi fL^4  \label{foeight}
\end{equation}
and we have shown the reason for its close resemblance to $Z$. When 
\begin{equation}
h=x^{n_1+1},\qquad f=x^{n_2+1}  \label{fonine}
\end{equation}
Eq (47) has five different analytical solutions for $\left( n_1,n_2\right)
=\left( 0,0\right) ,$ $\left( -5/2,-3\right) ,$ $\left( -14/5,-18/5\right) ,$
$\left( -11/5,-12/5\right) ,$ $\left( -5,-5\right) $ \cite{fitwo}. The
density and the pressures are given by the general formulae plus the
additions from Eq (44). In total, charged perfect fluid is a particular case
of the anisotropic fluid model.

One can find new neutral anisotropic solutions by prescribing the form of
the anisotropy factor. For example 
\begin{equation}
\Delta p=f\left( x\right) L^a,\qquad L_{xx}=\frac{3h}{4x}L^2+\frac{2\pi f}{%
\left( a-3\right) x}L^{a-1},  \label{fifty}
\end{equation}
where $a$ is some constant, not equal to $3$.

\section{Solutions with equation of state}

There is a case when the integration of Eq (47) is straightforward. Let the $%
x$-functions on the r.h.s become constants $\alpha ,\beta $. Multiplication
with $L_x$ gives a first integral 
\begin{equation}
L_x^2=\xi +\frac 23\alpha L^3+\frac 12\beta L^4  \label{fione}
\end{equation}
where $\xi $ is another constant. Integrating once more one finds 
\begin{equation}
\int \frac{dL}{\sqrt{\frac \beta 2L^4+\frac{2\alpha }3L^3+\xi }}=x+t\equiv u,
\label{fitwo}
\end{equation}
where an arbitrary function of time has been set to $t$. This is a
particular form of the integral 
\begin{equation}
u=\int_0^Lf\left( s\right) ^{-1/2}ds,\qquad f\left( s\right)
=a_0s^4+4a_1s^3+6a_2s^2+4a_3s+a_4  \label{fithree}
\end{equation}
and other examples of it will appear later. It may be inverted in terms of
the Weierstrass function $P\left( u,g_2,g_3\right) $ \cite{fithree}, namely 
\begin{equation}
L\left( u\right) =\frac{\sqrt{a_4}P_u+2a_3\left( P-\frac{a_2}2\right) +a_1a_4%
}{2\left( P-\frac{a_2}2\right) ^2-\frac 12a_0a_4},  \label{fifour}
\end{equation}
\begin{equation}
g_2=a_0a_4-4a_1a_3+3a_2^2,  \label{fifive}
\end{equation}
\begin{equation}
g_3=a_0a_2a_4+2a_1a_2a_3-a_2^3-a_0a_3^2-a_1^2a_4.  \label{fisix}
\end{equation}
Here $a_i$ are constants or functions of time. In the present case 
\begin{equation}
g_2=\frac 12\beta \xi ,\qquad g_3=-\frac 1{36}\alpha ^2\xi .  \label{fiseven}
\end{equation}
Eq (35) shows that while $L=L\left( u\right) $ the density in general is $%
\rho \left( t,r\right) $ because of terms with $x$. However, Eqs (44,45)
lead to their cancellation and tuning the expansion yields 
\begin{equation}
8\pi \rho \left( u\right) =3\left( C-4\xi u\right) +12LL_u,\qquad \Theta
^2=9C-36\xi t,  \label{fieight}
\end{equation}
where $C$ is a constant. Eq (19) holds for the effective energy and
pressures of the charged perfect fluid. It gives the following equation for
the sum of the genuine energy density and isotropic pressure 
\begin{equation}
8\pi \left( \rho +p\right) =\frac{4L^4}{L_u}\left( \frac 53\alpha +\frac
32\beta L\right) .  \label{finine}
\end{equation}
Hence, $p=p\left( u\right) $ and $\rho =\rho \left( p\right) $. The fluid
possesses an EOS and this is the charged Wyman solution \cite{twthree}, \cite
{twfour}, \cite{twfive}, \cite{twsix}. Combining the two previous equations
we obtain 
\begin{equation}
8\pi \left( \rho +6p\right) =-15\left( C-4\xi u\right) -60\xi \frac
L{L_u}+6\beta \frac{L^5}{L_u}.  \label{sixty}
\end{equation}
When the fluid is neutral ($\xi =0,\beta =0$) the EOS becomes linear and the
Wyman solution emerges \cite{three}, \cite{twtwo}.

In principle, the general solution of the shear-free anisotropic fluid
depends on an arbitrary function $L(t,r)$. Going to a perfect or charged
perfect fluid means imposing a differential equation on $L$ so that the
freedom reduces to functions of $t$ or $r$. Imposing further an EOS makes
the system overdetermined and solutions (if any) may exist only in some
exceptional cases, like Eq (60), which holds for constant $\xi ,\alpha
,\beta $.

Is it possible to generalize this EOS mechanism to anisotropy which is not
the result of charging the fluid? For this purpose we take 
\begin{equation}
Z=\frac 43xLF\left( L\right) _L,\qquad L_{xx}=F\left( L\right) _L,\qquad
L_x^2=2F\left( L\right) +g\left( t\right) ,  \label{sione}
\end{equation}
where $F\left( L\right) $ is an arbitrary function of $L$. The main equation
still may be integrated and $L=L\left( u\right) $ if the integration
function of time is constant $g_0$. Eq (35) gives 
\begin{equation}
4\pi \rho =\frac{\Theta ^2}6+6LL_u+2x\left( 2LF_L-6F-3g_0\right) .
\label{sitwo}
\end{equation}
The elimination of the $x$-term requires an equation for $F$ leading to the
Wyman solution. Thus, it is not possible to generalize it to anisotropic
fluids except for a charged perfect fluid.

\section{Conformally flat solutions}

These solutions require $\Psi _2=0$. The definition of $Z$ and the vanishing
shear constraint (37) then yield 
\begin{equation}
\Delta p=f\left( x\right) L,\qquad Z=-\frac{4\pi }3f\left( x\right) L,\qquad
L_{xx}=-\frac{\pi f\left( x\right) }x,  \label{sithree}
\end{equation}
$f\left( x\right) $ being an arbitrary function. The differential equation
is easily integrated so that 
\begin{equation}
L=G\left( x\right) +h\left( t\right) x+q\left( t\right) ,\qquad f\left(
x\right) =-\frac 1\pi xG\left( x\right) _{xx}  \label{sifour}
\end{equation}
and $L$ depends on three arbitrary functions. The expressions for $m,\rho
,p_r,p_t$ are rather cumbersome and are not given. Let us note a general
equation which follows from Eqs (18,40) 
\begin{equation}
2p_t+p_r+3\rho =\frac L{\dot L}\dot \rho .  \label{sifive}
\end{equation}

Things simplify considerably for a conformally flat perfect fluid. Then one
can put $G=0$ and Eqs (15, 35, 64) give 
\begin{equation}
4\pi \rho =\frac{\Theta \left( t\right) ^2}6+6h\left( t\right) q\left(
t\right) ,\qquad m=\frac{4\pi r^3\rho }{3L^3},  \label{sisix}
\end{equation}
so that $\rho $ is a function of time only. Eq (65) then gives for $p$%
\begin{equation}
p=-\rho +\frac{\dot \rho L}{3\dot L}.  \label{siseven}
\end{equation}
This is exactly the Stephani solution \cite{one}, \cite{thisix}. Eq (29)
shows that when $\Psi _2=0=\Delta p$ the shear vanishes identically. Thus
Stephani's solution describes all conformally flat perfect fluids, while the
solution (64) describes all conformally flat anisotropic fluids with
vanishing shear. Comparison between Eq (45) and Eq (63) shows that the
anisotropy can not be caused by charging the fluid unless the solution is
static, hence, evolving charged perfect fluid can not be conformally flat in
addition. This result was found in the past by other methods \cite{thifive}.

\section{Uniform density}

In the previous section we have seen that some conformally flat solutions
possess uniform density $\rho \left( t\right) $. One can find all solutions
of this kind. Up to now we have worked in the $L$-formalism, taking $L$ as a
basic function. From now on we shall work in the $Y$-formalism. For uniform
energy density Eq (17) may be integrated 
\begin{equation}
m=\frac{4\pi }3\rho Y^3-f\left( t\right)  \label{sieight}
\end{equation}
and Eqs (27, 38) give 
\begin{equation}
Z=\frac{f\left( t\right) }{Y^3},\qquad \frac{4\pi }3\Delta p=-\frac{\dot f}{%
2Y^2\dot Y}.  \label{sinine}
\end{equation}
The combination of Eqs (34, 68) provides a first order differential equation
for $Y$%
\begin{equation}
Y_z^2=AY^4+Y^2+2fY,  \label{seventy}
\end{equation}
\begin{equation}
A\left( t\right) =\frac{\Theta ^2}9-\frac{8\pi }3\rho ,\qquad z=\ln r,
\label{seone}
\end{equation}
which can be integrated 
\begin{equation}
\int \frac{dX}{\sqrt{2fX^3+X^2+A}}=s,\qquad X=\frac 1Y,\qquad s=\ln rB\left(
t\right) .  \label{setwo}
\end{equation}
Here another arbitrary function of time appears. One can invert the
functional dependence with the help of formula (54) and express $Y$ in terms
of the Weierstrass function with 
\begin{equation}
g_2=\frac 1{12},\qquad g_3=-\frac 1{216}-\frac{f^2A}4.  \label{sethree}
\end{equation}
The pressures are obtained from Eqs (65, 69).

In the case of perfect fluid the function $f\left( t\right) $ becomes a
constant and the above solution still applies. When this constant vanishes, $%
Z=0=\Psi _2$ and the solution is conformally flat. The integral above is
expressed in elementary functions. We find 
\begin{equation}
rX=L=e^{-\lambda }=-\frac AB+\frac B4r^2,  \label{sefour}
\end{equation}
which coincides with Eq (64) when $G=0$. These results are the same as the
classical results on perfect fluids with uniform density \cite{one}, \cite
{twseven}. They can be rewritten in the $L$-formalism. There $L$ satisfies
Eq (42) with $F\left( x\right) \sim x^{-5/2}$ which follows from Eq (69).

In the general anisotropic case one has $Y=Y\left( s\right) $ when $f$ and $%
A $ are constants. Then $\Delta p=0$ and the fluid is in fact perfect.
Further, $m=m\left( s\right) $ when $\Theta $ is constant. This leads to the
constancy of $\rho $ and $p.$ Then Eqs (18, 68) show that $\rho +p=0$. Once
again the fluid has an EOS when the arbitrary time functions become
constants.

\section{Solutions with $\rho +p_r=0$}

We may generalize the uniform density case with the help of an arbitrary
function $F\left( Y\right) $. Mimicking Eq (69), let us take 
\begin{equation}
Z=f\left( t\right) F\left( Y\right) _YY.  \label{sefive}
\end{equation}
In particular, uniform density follows from $F=-1/3Y^3$. Eq (27) gives the
mass function 
\begin{equation}
\frac m{Y^3}=3f\left( t\right) F\left( Y\right) +g\left( t\right)
\label{sesix}
\end{equation}
and then $\rho $ is found from Eq (25) 
\begin{equation}
\frac{4\pi }3\rho =f\left( 3F+YF_Y\right) +g.  \label{seseven}
\end{equation}
Proceeding like in the previous section we obtain 
\begin{equation}
\int \frac{dX}{\sqrt{-6fF\left( \frac 1X\right) +X^2+A_1}}=s,\qquad
A_1\left( t\right) =\frac{\Theta ^2}9-2g,  \label{seeight}
\end{equation}
which determines $Y\left( s,t\right) $. We also find 
\begin{equation}
4\pi \left( \rho +p_r\right) =-\left( 3\dot fF+\dot g\right) \frac Y{\dot
Y},\qquad -\frac{8\pi }3\Delta p=\dot fF_Y\frac{Y^2}{\dot Y}+4Z,
\label{senine}
\end{equation}
which determine the pressures. When $f\left( t\right) ,g\left( t\right) $
are constants we get a linear EOS $\rho +p_r=0$. The density and pressures
are not constant like in the previous section.

The appearance of EOS can be seen in another way too. We have $m=m\left(
Y\right) $ from Eq (76). Eqs (17,18) indicate that in this case the
difference between time and space derivatives effectively disappears and $%
\rho +p_r=0$. The discussion in the present section is analogous to that in
Sec. 6 which lead to the Wyman solution. However, now we work in the $Y$%
-formalism and the EOS differs from Eq (60), reached in the $L$-formalism.

\section{Solutions with $p_t=\gamma p_r$ or $p_r=0$}

Solutions with vanishing tangential or radial pressure have been discussed
in the past for time-dependent anisotropic fluids with shear \cite{fifour}, 
\cite{fifive}, \cite{fisix}. We shall investigate the more general case of
linear EOS between the pressures 
\begin{equation}
p_t=\gamma p_r.  \label{eighty}
\end{equation}
Here $\gamma $ is some constant measuring the degree of anisotropy between
the two extreme cases $\gamma =0$ ($p_t=0$) and $\gamma =1$ ($p_t=p_r$,
isotropy condition). We have shown that for shear-free fluids $p_t$
satisfies Eq (40), similar in structure to Eq (18) for $p_r$. This allows to
write Eq (80) as a time derivative and to integrate it, obtaining 
\begin{equation}
\frac 32Y^3Z+\left( 1-\gamma \right) m=g\left( r\right) ,  \label{eione}
\end{equation}
$g\left( r\right) $ being arbitrary. In addition, $Z$ satisfies Eq (26) for
any spherically symmetric metric. This leads to 
\begin{equation}
\left( \frac m{Y^{1+2\gamma }}\right) ^{\prime }=\frac{2g\left( r\right)
Y^{\prime }}{Y^{2\left( 1+\gamma \right) }}.  \label{eitwo}
\end{equation}
Suppose that $g\left( r\right) =g_0$, a constant. Then another integration
gives 
\begin{equation}
m=a+k\left( t\right) Y^{1+2\gamma },\qquad a=-\frac{2g_0}{1+2\gamma }
\label{eithree}
\end{equation}
and $k\left( t\right) $ is another arbitrary function. Eqs (81, 17) read 
\begin{equation}
Z=-\frac a{Y^3}+\frac{2\left( \gamma -1\right) }3kY^{2\left( \gamma
-1\right) },  \label{eifour}
\end{equation}
\begin{equation}
4\pi \rho =\left( 1+2\gamma \right) kY^{2\left( \gamma -1\right) }.
\label{eifive}
\end{equation}
Eqs (38, 80) result in 
\begin{equation}
4\pi p_r=-4\pi \rho -\dot k\frac{Y^{2\gamma -1}}{\dot Y},  \label{eisix}
\end{equation}
which allows to find $\Psi _2$ from the definition of $Z$%
\begin{equation}
\Psi _2=-\frac a{Y^3}+\frac 13\left( \gamma -1\right) \left( 1-2\gamma
\right) kY^{2\left( \gamma -1\right) }-\frac 13\left( \gamma -1\right) \dot k%
\frac{Y^{2\gamma -1}}{\dot Y}.  \label{eiseven}
\end{equation}
The combination of Eqs (34, 83) provides a first-order equation for $Y$ or $%
X=1/Y$ which can be integrated 
\begin{equation}
\int \frac{dX}{\sqrt{-2aX^3+X^2-2kX^{2\left( 1-\gamma \right) }+\Theta ^2/9}}%
=s.  \label{eieight}
\end{equation}
The integral may be inverted with the Weierstrass function when the
polynomial $Q$ under the root is up to the fourth degree. This happens when $%
\gamma =0,\pm 1/2,\pm 1$.

In the case $\gamma =0$ ($p_t=0$) the polynomial and the other
characteristics of the fluid read 
\begin{equation}
Q=-2aX^3+\left( 1-2k\right) X^2+\frac{\Theta ^2}9,  \label{einine}
\end{equation}
\begin{equation}
m=a+kY,\qquad Z=-\frac a{Y^3}-\frac{2k}{3Y^2},  \label{ninety}
\end{equation}
\begin{equation}
4\pi \rho =\frac k{Y^2},\qquad 4\pi \left( \rho +p_r\right) =-\frac{\dot k}{%
Y\dot Y},  \label{nione}
\end{equation}
\begin{equation}
\Psi _2=-\frac a{Y^3}+\frac k{3Y^2}+\frac{\dot k}{3Y\dot Y}.  \label{nitwo}
\end{equation}
When $k$ is constant $\rho +p_r=0,$ $m=m\left( Y\right) $, $Z=Z\left(
Y\right) $ and the solution belongs also to the class discussed in the
previous section.

In the case $\gamma =1$ (perfect fluid) one finds 
\begin{equation}
Q=-2aX^3+X^2+\frac{\Theta ^2}9-2k,  \label{nithree}
\end{equation}
\begin{equation}
m=a+kY^3,\qquad Z=-\frac a{Y^3},  \label{nifour}
\end{equation}
\begin{equation}
4\pi \rho =3k,\qquad 4\pi \left( \rho +p_r\right) =-\frac{\dot kY}{\dot Y}%
,\qquad \Psi _2=-\frac a{Y^3}.  \label{nifive}
\end{equation}
Now $\rho =\rho \left( t\right) $ and this is in fact a uniform density
solution given by Eq (72) when $f\left( t\right) =-a.$ When $a=0$ the
solution is conformally flat as can readily be seen from Eq (95).

Let us discuss finally the case $p_r=0$. Eq (18) gives $m=m\left( z\right) $%
. Substituting this into Eq (34) yields 
\begin{equation}
X_z^2=-2m\left( z\right) X^3+X^2+\frac{\Theta ^2}9.  \label{nisix}
\end{equation}
This equation may be integrated easily when $m$ is constant and a formula
close to Eq (72) or to Eq (88) with $Q$ from (89) or (93) is obtained.
However, this is unphysical case because the energy density vanishes.
Therefore Eq (96) should be solved by other methods.

\section{Conclusions}

Recently we have stressed the importance of anisotropic fluid models for
astrophysics \cite{fofour}. With the present paper we start their systematic
study from the most general case. A decomposition of the mass function into
a local and global part is redireved and related to the second Weyl
invariant. It holds for any radiating anisotropic fluid which is spherically
symmetric. Then, for simplicity, we turn off the radiation and combine this
mass formula with the reformulation of the Einstein equations in terms of
the mass function (given by a classic formula) and its derivatives. This
results in the important relation (27) for the master potential $Z$ and
another expression for the shear, Eq (29). Here the usual metric formula for
the shear (23) is elevated to global and source quantities such as $\Psi _2$
and the anisotropy factor. It allows to push further the study of shear-free
fluids, providing an important constraint. Now everything can be expressed
in terms of $L$ and Eq (27) transforms into the second order Eq (36). The
general solution of this equation is based on one arbitrary function which
may be $Y$, $L$, $Z$, $\Delta p$ or $\Psi _2$. The treatment of the
tangential pressure becomes on an equal footing with the radial pressure due
to formula (40). The acceptance of maximum freedom in the system allows to
obtain an anisotropic shear-free core of simple relations. Only after that
we start to impose one by one different constraints aiming the system at
particular cases.

Thus perfect fluid is this core with imposed isotropy condition ($\Delta p=0$%
), charged fluid is a neutral fluid with a special kind of anisotropy,
conformal flatness means the constraint $\Psi _2=0$, uniform density
constrains the energy density to a function of time, anisotropy may be
studied in more detail by a linear equation among the pressures, which
includes the cases where one of them vanishes. In this way the relations
among the particular cases become much clearer due to the relations of all
of them to the core. Time and again second order equations like Eq (50)
appear, which sometimes may be integrated to first order, like Eq (61) and
even to implicit integral formulas for the solution, like Eqs (52, 72, 78,
88). These are different examples of the integral (53) which explains the
persistent appearance of the Weierstrass elliptic function. In many cases,
however, it reduces to elementary functions. This is true also for the $Y$%
-formalism.

Imposing one constraint fixes the arbitrary potential of the general model
but up to functions of time or radius. Imposing a second constraint makes
the system overdetermined, yet special solutions still exist when the
arbitrary functions become constants. Then the integral formulas reduce the
dependence on $t$ and $r$ to a dependence on a single variable $r^2+t$ or $%
rB\left( t\right) $. The result is appearance of equation of state, linear
in some cases, as seen from Eqs (60, 79).

In total, we find the general charged anisotropic solution, all conformally
flat solutions, all uniform density solutions, a large class of solutions
with proportional or vanishing pressures and solutions with vanishing sum of
density and radial pressure.

Finally, let us discuss the static case. Unfortunately, Eqs (16-21) are
incomplete then because (16,18,19) are trivial. The shear trivially vanishes
and formula (31) holds no more. The metric function $\nu $ is not determined
by $Y$ and becomes another free potential. One can use it and $\Delta p$ to
express all static spherically symmetric anisotropic solutions \cite{fiseven}%
.

\end{document}